\documentclass[12pt]{article}
\usepackage{times}
\usepackage{geometry}
\usepackage[utf8]{inputenc}
\usepackage{enumitem,amssymb}
\usepackage{ragged2e}
\newlist{thematic}{itemize}{8}
\setlist[thematic]{label=$\square$}
\usepackage{pifont}

\usepackage{graphicx}
\usepackage{epsfig}
\usepackage{color}
\usepackage{txfonts}
\usepackage{wrapfig}
\usepackage{sidecap}
\usepackage{natbib,twoopt}
\usepackage[bookmarks=false,colorlinks=true, citecolor=blue, backref=true,breaklinks=true]{hyperref}
\usepackage[font={scriptsize},labelfont={scriptsize,bf}]{caption}
\begin{document}

\raggedright
\huge
{\bf Astro2020 Science White Paper }\linebreak

{\LARGE \bf Warm H$_2$ as a probe of massive accretion and feedback through shocks and turbulence across cosmic time}\linebreak
\normalsize

\noindent \textbf{Thematic Areas:} \hspace*{60pt} $\square$ Planetary Systems \hspace*{10pt} $\square$ Star and Planet Formation \hspace*{20pt}\linebreak
$\square$ Formation and Evolution of Compact Objects \hspace*{31pt} $\boxtimes$ Cosmology and Fundamental Physics \linebreak
  $\square$  Stars and Stellar Evolution \hspace*{1pt} $\square$ Resolved Stellar Populations and their Environments \hspace*{40pt} \linebreak
  $\boxtimes$    Galaxy Evolution   \hspace*{45pt} $\square$             Multi-Messenger Astronomy and Astrophysics \hspace*{65pt} \linebreak
  
\vspace{-0.2 cm}
\textbf{Principal Author:}

Name: Philip N. {\bf Appleton}	
 \linebreak						
Institution: Caltech 
 \linebreak
Email: apple@ipac.caltech.edu
 \linebreak
Phone: (+1) 626 780 6358 
 \linebreak
\textbf{Co-authors:}
\linebreak
Lee {\bf Armus}: Caltech; 
Francois {\bf Boulanger}: Observatoire de Paris; 
Matt {\bf Bradford}: JPL; 
Jonathan {\bf Braine}: University of Bordeaux; 
Volker {\bf Bromm}: University of Texas at Austin; 
 Peter {\bf Capak}: Caltech; 
 Michelle {\bf Cluver}: Swinburne University of Technology;
Asantha {\bf Cooray}: UC Irvine;
Tanio {\bf Diaz-Santos}: Universidad Diego Portales; 
Eiichi {\bf Egami}: University of Arizona;
Bjorn {\bf Emonts}, NRAO, Challottesville;
Pierre {\bf Guillard}: IAP-Paris; 
George {\bf Helou}: Caltech; 
 Lauranne {\bf Lanz}: Dartmouth College; 
Susanne {\bf Madden}: CEA, Saclay; 
Anne {\bf Medling}: University of Toledo; 
Ewan {\bf O'Sullivan}: Center for Astrophysics $\vert$ Harvard-Smithsonian;
Patrick {\bf Ogle}: STScI; 
 Alexandra {\bf Pope}: University of Massachusetts; 
Guillaume {\bf Pineau des For\^ets}: IAS, Orsay; 
J. Michael {\bf Shull}: U. of Colorado; 
John-David {\bf Smith}: University of Toledo; 
Aditya {\bf Togi}: University of St. Marys;
C. Kevin {\bf Xu}: National Astronomical Observatories, Beijing

\vspace{-0.15 cm}
\justify 
%In the next decades, huge gains in studying warm H$_2$ are achievable with future instrumentation that will unlock many of the mysteries of galaxy formation and evolution.
Galaxy formation depends on a complex interplay between gravitational collapse, gas accretion, merging, and feedback processes. Yet, after many decades of investigation, these concepts are poorly understood. This paper presents the argument that warm H$_2$ can be used as a tool to unlock some of these mysteries.   Turbulence, shocks and outflows, driven by  star formation, AGN activity or inflows, may prevent  the rapid buildup of star formation in galaxies. Central to our understanding of how gas is converted into stars is the process by which gas can dissipate its mechanical energy through turbulence and shocks in order to cool. H$_2$ lines provide direct quantitative measurements of  kinetic energy dissipation in molecular gas in galaxies throughout the Universe. Based on the detection of very powerful H$_2$ lines from z = 2 galaxies and proto-clusters at the detection limits of {\it Spitzer}, we are confident that future far-IR and UV H$_2$ observations will provide a wealth of new information and insight into galaxy evolution to high-z.  Finally, at the very earliest epoch of star and galaxy formation, warm H$_2$ may also provide a unique glimpse of molecular gas collapse at 7 $<$ z $<$ 12 in massive dark matter (DM) halos on their way to forming the very first galaxies. Such measurements are beyond the reach of existing and planned observatories.  
\pagebreak

\vspace{0.3 cm}
\noindent
{\bf The importance of direct H$_2$ detection of warm gas throughout the universe}. H$_2$ is an important coolant of shocked gas. Pure rotational (quadrupole) transitions radiate in the rest-frame mid-IR, and were routinely detected in the spectra of nearby galaxies by {\it ISO} and {\it Spitzer}.   Unlike indirect measurements of molecular hydrogen mass through uncertain conversion from trace molecules such as  CO, the {\it direct detection} of emission lines from molecular hydrogen (H$_2$) is easily achieved with cryogenically-cooled space telescopes when the gas is heated above $\sim$100K, or fluorescently excited (\citealt{shu82}). Alternatively, H$_2$ can be detected in absorption in the far-UV through the Lyman and Werner-bands (e.g., \citealt{rac09}).  
 
When star formation dominates galaxies, photo-electrons,  ejected from small grains and PAH molecules by UV photons, drive the heating of the neutral gas (Fig. 1a). However,  non-radiative forms of excitation of the H$_2$ are also present, such as gas heated in shocks and turbulence in subsets of galaxies, collisional systems or groups (e.g., Fig. 1b; \citealt{app06,rou07,clu10,pet12, clu13,sti14}), radio galaxies (\citealt{ogl07, ogl10,gui12a}), and cluster galaxies (\citealt{ega06,ogl12,siv14}). Most  show unusually strong H$_2$ line-luminosities, high H$_2$/FIR and H$_2$/PAH ratios, often in regions devoid of obvious star formation. Cosmic rays have also been suggested as another source of heating in some galaxy cluster sources (\citealt{fer08}), as well as X-rays produced by accreting super-massive black holes (Active Galactic Nuclei - AGN).  Observations and modeling of a well studied nearby example (Fig. 2a) suggest that low-velocity magnetohydrodynamic (MHD) shocks are responsible for the high power in the lowest-lying, most luminous, rotational lines (\citealt{gui09, les13, app17}) which encode information about the most massive H$_2$ component. 

JWST will provide a remarkable view of some of the brightest IR H$_2$ lines in the nearby (z $<$ 2) universe, but its wavelength coverage is limited.  However, the next generation of wide-band Far-IR (FIR) and Far-UV (FUV) telescopes will open a new window into the turbulent universe to cosmological distances,  including the virialized gas that cools to form early galaxies.  

\begin{figure}[h]
\vspace{-0.3cm}
\includegraphics[width=0.50\textwidth]{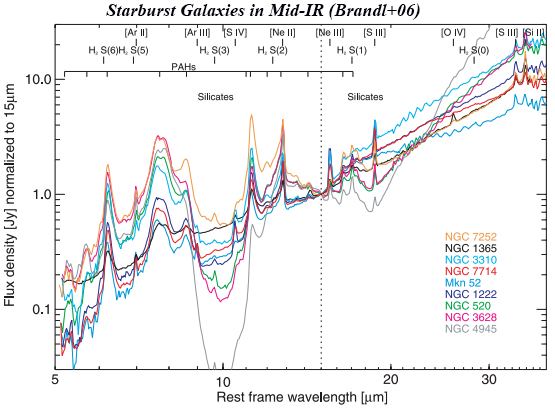}
\includegraphics[width=0.50\textwidth]{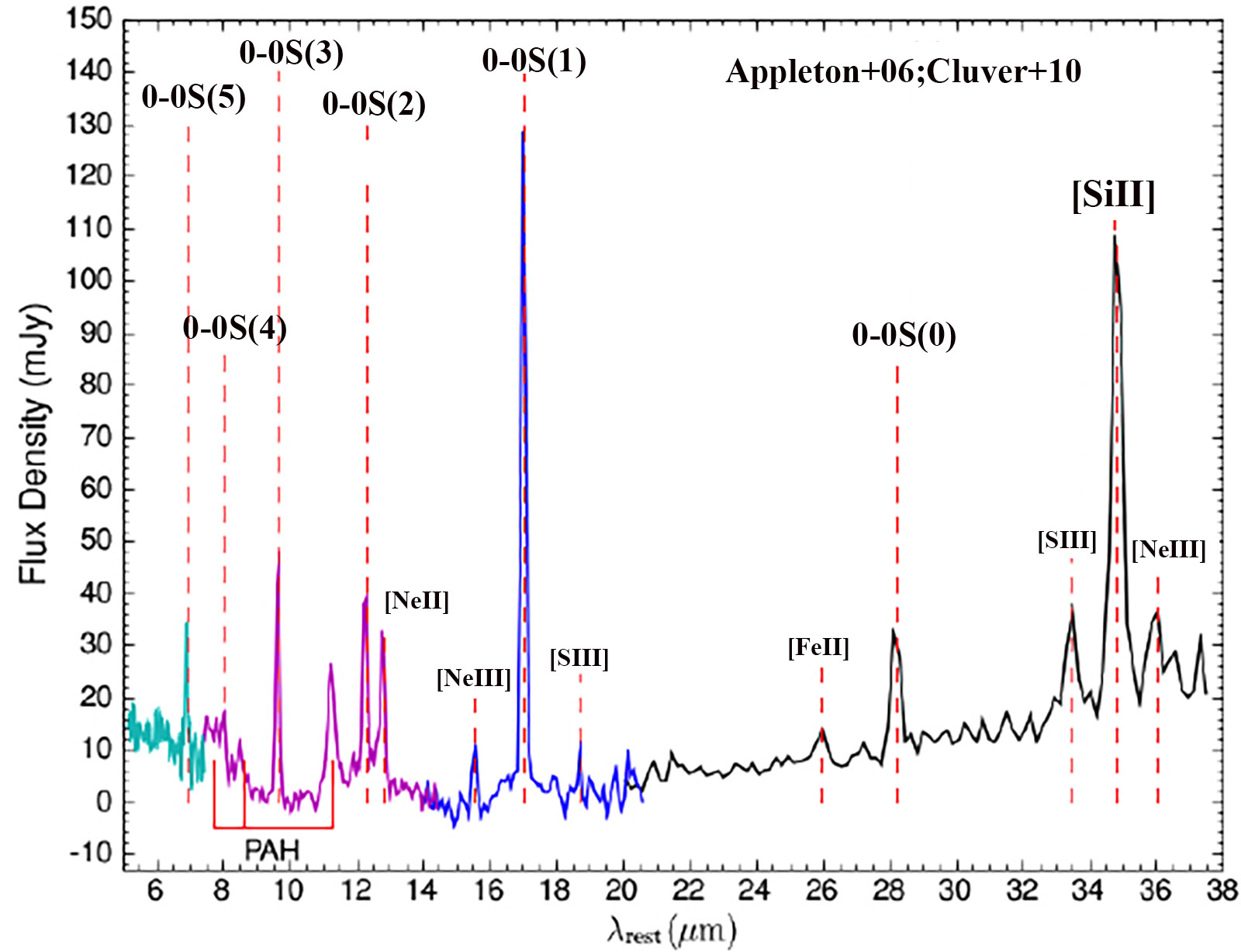} 
\vspace{-0.2 cm}
\caption{{\it Spitzer} spectra showing the contrast between galaxies excited by star formation processes compared with shocks (very strong H$_2$ lines) }
\label{fig:fig1}
\end{figure}

\begin{figure}[h]
\includegraphics[width=0.98\textwidth]{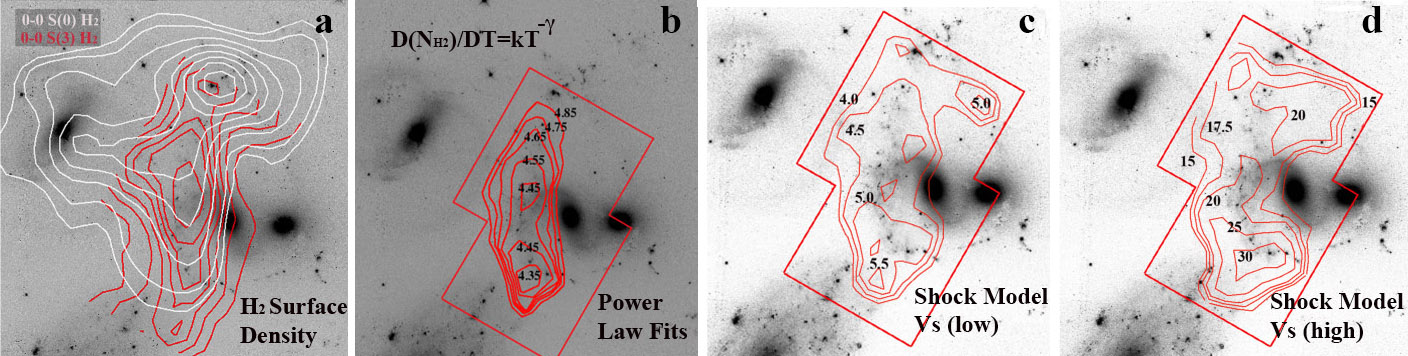}
\caption{An example of the power of spectral mapping of many warm H$_2$ emission lines in the Stephan's Quintet intergalactic filament (\citealt{app17}) showing, (a) contours of H$_2$ surface density in the 0-0 S(0)28$\mu$m and 0-0 S(1)17$\mu$m lines, (b) distribution of power-law excitation index \citep{tog16} showing unusually low values (higher T) in the shock and (c,d) distribution of MHD modeled molecular shock velocities. }
\label{fig:fig2}
\vspace{-0.5 cm}
\end{figure}

\vspace{-0.4 cm}

H$_2$ excitation diagrams are a powerful way of exploring the distribution of level populations in H$_2$ molecules in galaxies (\citealt{rig02,rou07}) and are especially powerful when many H$_2$ transitions are measured together.  In galaxies, multiple temperature components are often present, resulting in a distribution of curved points in the excitation diagrams.  This is most naturally explained by a power law distribution of temperatures (\citealt{tog16}).  In systems suspected of shock or turbulent heating, a large fraction (perhaps all) of the H$_2$ is warm (T $>$ 200 K) and exhibits low values of power-law index (Fig. 2b) and can be fit with shock models to derive important properties (Fig. 2c,d), such as mass flow into the shock and the total energy dissipation. Observations of the intergroup gas in Stephan's Quintet reveal that the bulk of the energy deposited on large scales is involved in a turbulent cascade and is dissipated mainly through H$_2$ line emission (\citealt{gui12b,app13, app17}), with a second component from Ly$\alpha$ emission (Guillard et al.  in prep). The cooling time of the gas through these lines is so short that kinetic energy must be injected faster than it can dissipate through line cooling. Strong mechanical forcing creates supersonic turbulence in the H$_2$ gas (\citealt{gui09}) and leads to powerful H$_2$ line emission as the dominant cooling channel (\citealt{app13,app17}). In these systems, rotational H$_2$ line emission  provides a direct measure of the turbulent dissipation rate. 

\vspace{0.1cm}
\noindent
{\bf  H$_2$ as a Tracer of Gas Accretion in Galaxies, Clusters and ProtoClusters }: 

The details of how gas accretes onto galaxies, and how feedback from star formation and AGNs might redistribute, heat and stir-up the gas, is still one of the unsolved problems in galaxy evolution.  In the standard ``hot mode" $\lambda$CDM picture, gas accumulating in DM halos above a critical  mass (M $>$ 10$^{12}$ M$_{\odot}$) is shock-heated at the virial radius (\citealt{ree77}), but then cools, eventually raining down isotropically to form a slowly rotating galaxy. Shocked molecular hydrogen lines would be detected just inside the main virial shock as the gas passes through the warm stage. Star formation and feedback from %possible 
AGNs will also drive further heating. 
Alternatively, at lower halo masses, gas can accumulate at the center through ``cold flows" of diffuse, T = 10$^4$ K gas (\citealt{far01,dek06}). The gas can flow into the central forming galaxy without developing strong shocks, creating high angular momentum proto-disks, with spiral-like filamentary morphologies (\citealt{dan15,ste17}). In such cases, rapid cooling by shocks is likely to occur close to the center of the galaxy, and major cooling to form stars may be more centrally concentrated. Feedback from AGN (e.g., radio jets) can also complicate the picture (\citealt{rus14,fau16,cor18}). 

%LATER??With enough information, combining radio observations of active jets, observations of the dynamics of the trace molecules with ALMA and the NGVLA, and state of the art excitation modeling, the powerful rotational H$_2$ lines will provide vital new insight into galaxy formation.  

\begin{figure}[h]
\vspace{-0.3 cm}
\includegraphics[width=0.35\textwidth]{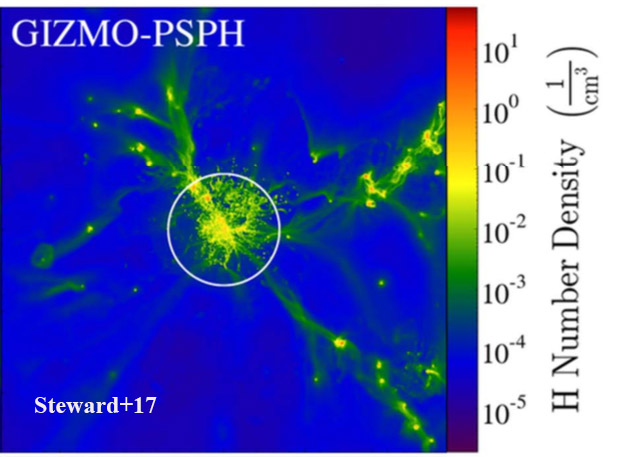}
\includegraphics[width=0.23\textwidth]{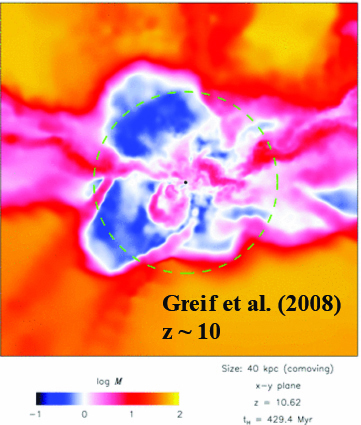}
\includegraphics[width=0.35\textwidth]{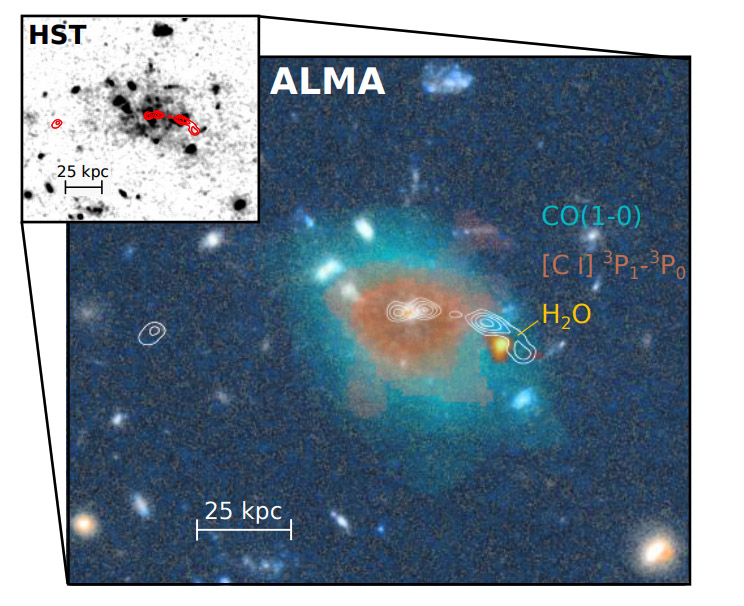}
\caption{ \scriptsize{(Left) Cold mode filaments extending into the center of a DM halo at z = 3 from the virial radius (circle), (middle) hot mode shocked gas shows pile-up inside the virial radius, and (right) the huge CO envelope associated with the Spiderweb protocluster, showing [CI] and H$_2$O emission \citep{emo18a}. This protogalaxy contains the brightest rotational H$_2$ line ever detected in any object by {\it Spitzer} \citep{ogl12}. }
}
\label{fig:fig3}
\vspace{-0.5 cm}
\end{figure}

\noindent
{\bf Exploring the Warm H$_2$ properties of Galaxies across Mass Scales and Redshift:}  

There are already strong hints that the the H$_2$ rotational lines will be powerful in galaxies at high z. \citet{fio10} discovered unusually strong 0-0S(3)9.6$\mu$m emission in the stacked {\it Spitzer} spectra of 16 z$\sim$2 ULIRGs, suggesting that prominent rotational lines will be common in IR luminous galaxies at z $\sim$2-3. 
Even more exciting was the discovery of tremendous luminosity (3 $\times~10^{10} L_{\odot}$) in the single observed H$_2$$\lambda$9.66$\mu$m line detected in center of the massive Spiderweb proto-cluster  at z = 2.15 (\citealt{ogl12}). Unfortunately, {\it Spitzer} was unable to measure more than one line because of limited spectral coverage and sensitivity, and so a complete picture of the extent, luminosity and excitation properties of the warm gas awaits future FIR observations. The core of the Spiderweb protocluster contains a vast extended molecular reservoir emitting in CO and [CI] \citep{emo16,emo18b}, with H$_2$O observations implying that shock heating and localized gas cooling occurs along the powerful radio jet (\citealt{gul16}; Fig. 3c).  It is thus possible that a significant fraction of the H$_2$ luminosity comes from a combination of  accretion power and/or gas heated in the radio jet. Radio jets can deposit significant power which emerges as H$_2$ emission (\citealt{ogl10,lan15,lan16}).  

{\it A key to exploring the H$_2$ excitation properties over a wide range of redshifts in the FIR is broad spectral coverage} allowing the capture of many transitions in the same spectrum, enabling estimates of the gas temperatures and masses. The often powerful H$_2$ 0-0S(3) and S(1) lines are unobservable above a redshift of 1.7 by JWST/MIRI. 

\vspace{0.2 cm}
\noindent
Future deep, large-volume, surveys of the Universe with sensitive FIR and FUV  spectrometers in the next generation of space telescopes will enable 
the  following science goals:

\vspace{0.1 cm}
\noindent
{\bf $\bullet$ Characterizing the full H$_2$ excitation properties (temperatures, warm H$_2$ masses) of  large samples of galaxies spanning critical halo masses in the range 10$^{11}$ to 10$^{13}$ M$_{\odot}$}. Using well developed modeling tools for photodissociation regions and shocks, combined with observations of other species, such as HD, [CI], CO, and far-IR [OI] and [CII] emission, it will be possible to explore those cases where turbulence and shocks dominate over star formation. Trends with mass, environment, redshift (cold mode accretion should become more common at higher z as halos become less massive), and metallicity will be explored.  Since almost nothing is known about warm H$_2$ above z = 0.5, this will open up completely unexplored territory. 

\noindent
{\bf $\bullet$ Spatial mapping of warm H$_2$ on scales commensurate with the virial radius in multiple H$_2$ rotational lines for galaxies at z = 2-3}, at the peak in the cosmic star formation rate energy density. Mapping the full spectral properties of the lowest rotational H$_2$ transitions inside the virial radii of galaxies at 2 $<$z $<3$ will provide a strong constraint on the most massive, cooler, gas reservoirs. They will span a range of halo masses where gas infall transitions from hot mode to cold flow accretion.  At z = 3, for a galaxy with a halo mass of few x 10$^{12}$ M$_{\odot}$,  the typical virial radius at which infalling gas would  shock and undergo rapid cooling in the hot mode accretion model is $\sim$100-140 kpc (12-18$"$). The 0-0S(1)17$\mu$m line (shifted to 70$\mu$m) is likely to be the strongest line in the typical T = 150-300 K range. Future FIR telescopes with apertures $>$ 3\,m could resolve gas collapsing from the virial radius to the center. Lower halo mass galaxies with ``cool flows" will likely exhibit different warm H$_2$ morphologies than the hot-mode examples, since the gas can smoothly fall directly to the center before undergoing significant compression.  Multi-beam mapping in many rotational H$_2$ lines simultaneously (as envisaged in e.g. the OST concept) will be incredibly powerful for examining the sources of heating and multi-waveband morphologies in galaxies both near and far. 

\noindent
{\bf $\bullet$ Identifying the role of AGN-radio mode feedback in redistributing the gas both spatially and into different phases}.  Warm H$_2$ measurements, NIR/FIR AGN line diagnostics and IR SEDs, when combined with higher resolution radio continuum (SKA, ngVLA) and  other molecular species (ALMA),  will help us disentangle the complexity of AGN interactions with their surroundings (\citealt{u19}). AGN jets are known locally to heat the gas and
suppress star formation (e.g.~\citealt{ogl12}; Ogle 2019 in prep.)  where the warm H$_2$ dominates the mass in the outflows (unlike CO and HCN which are often just tracers). H$_2$ diagnostics may also be helpful in measuring the large-scale reservoirs of gas heated by the AGN. Such gas may be a critical link to an AGN feedback cycle, which is central to structure formation and galaxy evolution models. Extensive filaments of ionized and molecular gas at the center of lower-z cool-core clusters and groups likely fuel the AGN.   Based on {\it Spitzer} observations, such clusters are known as powerful warm H$_2$ emitters.  However, the cooling mechanism in the filaments is poorly understood, and it is not certain what triggers the gas to precipitate out of the hot phase, nor how it cools so rapidly. H$_2$ detection will be an excellent new window into the AGN heating and cooling cycle as observations progress to higher and higher redshift, where the balance between different kinds of AGN feedback (radiatively efficient versus jet-dominated) may shift. 
 
\noindent
{\bf $\bullet$ Detecting H$_2$ in cold flows through UV absorption} Although we expect to trace cold molecular filaments as far as possible in emission in the molecule hydrogen lines and other mid-IR atomic lines, the detection of Lyman and Werner-band H$_2$ in absorption along the line of sight to AGN has previously been demonstrated (\citealt{sav77,tum02,rac09}), and will be feasible to z = 3 with the next generation of large FUV/UVO telescopes (e.g., LUVOIR). This will enable very sensitive measurements of a possible molecular component to cold filaments along sight lines that pass near large mass concentrations.  Although cold flows are expected to be mainly ionized by background UV radiation, merging substructures, flow instabilities and shocks may create pockets of higher density gas that might lead to {\it in situ} star formation within the cold flows. Such stars will populate the outer virialized galaxy halos with time.  Higher density clumps are common features of numerical models of the cosmic web (Fig. 3a). 

\noindent
{\bf $\bullet$ Searching for H$_2$ in massive DM halos at z $>$ 7. }  In very low-metallicity environments, gas cooling in massive galaxies will be dominated by Ly$\alpha$ and H$_2$ pure-rotational lines.  Since Ly$\alpha$ emission can be easily extinguished via absorption and scattering in the increasingly neutral intergalactic medium (IGM) at high z, H$_2$ emission lines will serve as a powerful probe for such first-generation galaxies.  Estimating H$_2$ line luminosities in early galaxies has been a focus of theoretical studies for many years (\citealt{omu03,san06,miz05,gon13,liu19}). Such predictions contain many uncertainties, such as the importance of self-shielding effects from dissociative UV radiation from the first stars (\citealt{wis08}), the growth (%either
with or without grains) and destruction of H$_2$ in shocks, as well as other feedback effects. Magnetic fields may also play role in reducing compression via C-shocks, lowering the gas temperature, and boosting rotational $H_2$ emission (see \citealt{les13}). 

Most predictions suggest\footnote{except for very rare DM halo masses in excess of 10$^{12}$ M$_{\odot}$ (\citealt{liu19})} that the direct detection of H$_2$  line luminosities for realistic halos would be challenging, but not impossible, for future cold, large IR telescopes like OST, especially with gravitational lensing. %if gravitational lensing is available.
For example, utilizing strong lensing clusters to provide lensing amplifications factors of 6-10 for H$_2$ point-like cores, we have identified a range of DM halo masses (Fig. 4a) that are common enough at z = 8 to fall within the lensing volume of a small sample of bright lensing clusters. In such cases, lensing can potentially provide a strong enough H$_2$ signal to be detectable with currently envisaged large cold FIR sensitive telescopes (Fig. 4b). Finally, we emphasize that the proposed deep FIR spectroscopic observations will also naturally address the onset of heavy elements. Small amounts of metal enrichment can shift the cooling from H$_2$ to metal fine-structure lines (e.g., [Si II]34.8$\mu$m, [Fe II]25.99/35.35$\mu$m; [CII]158$\mu$m; \citealt{san06}), many of which are also observable in the FIR  at high-z.
\begin{figure}[ht]
\centering
\vspace{-0.2 cm}
\includegraphics[width=0.4\textwidth]{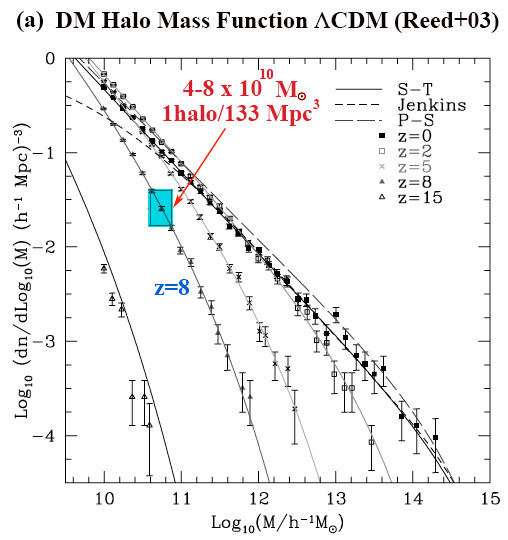}
\includegraphics[width=0.5\textwidth]{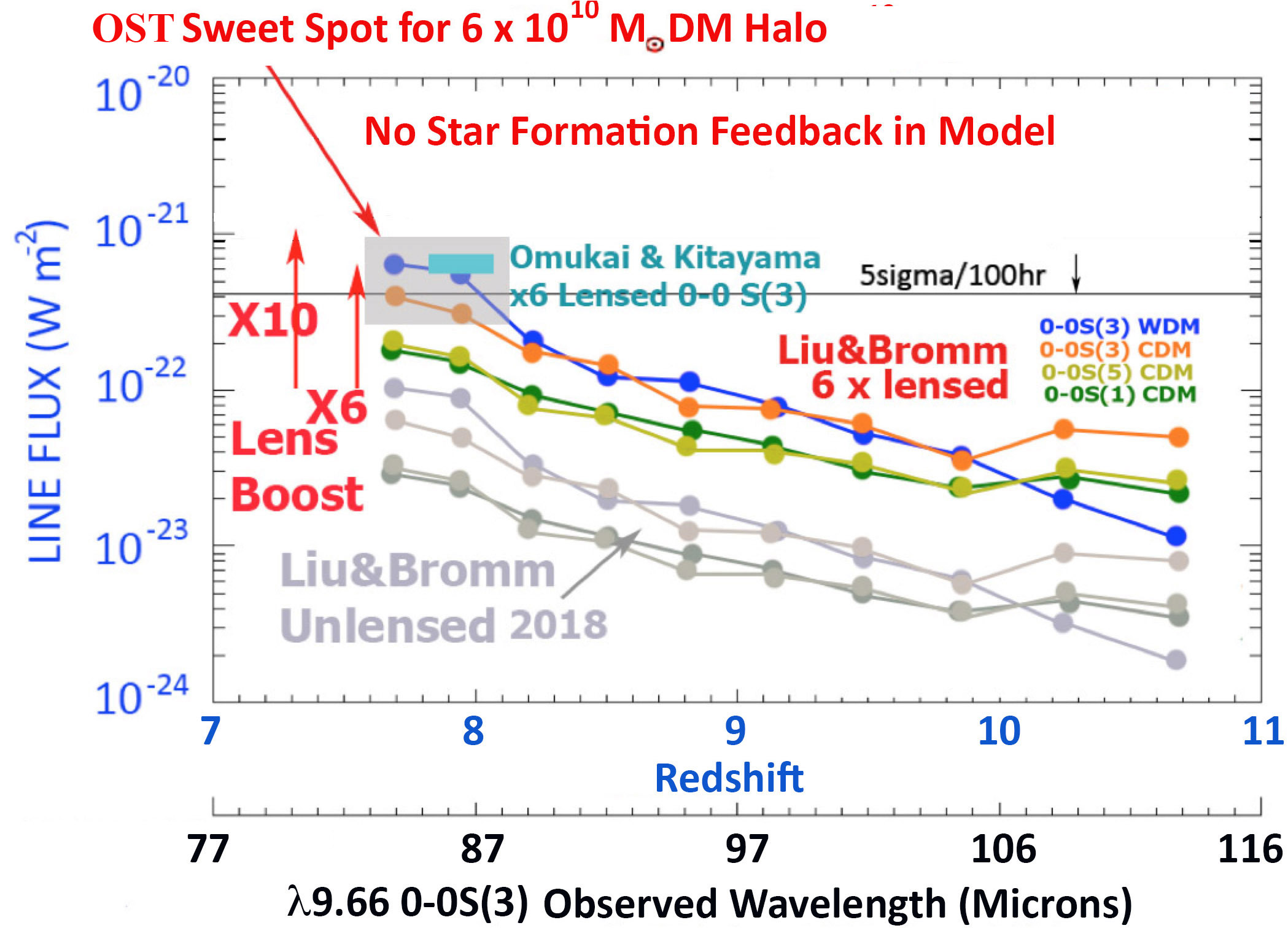}
\caption{(Left) Dark matter mass function for $\lambda$-CDM showing the target mass range that provides (at z = 8) approximately 1 halo per 133 Mpc$^3$. This is $\sim$ the co-moving  z = 8 sample volume if the lensing caustics of a single rich lensing cluster is observed. (Right) Predictions for 6 and 10 x lensed (colored lines) and un-lensed (grey lines) for H$_2$ line emission as a function of redshift for a halo of mass 6.9 x 10$^{10}$ M, extrapolated from the models of \citet{liu19} and \citet{omu03}. Also shown is the 5$\sigma$ sensitivity threshold for the Origins Space Telescope concept. Detection may be possible around z = 8 with lensing.}
\label{fig:fig4}
\vspace{-0.4 cm}
\end{figure}

\noindent
{\bf In summary: The direct detection of H$_2$ in the Universe is vital for the following reasons: }

 {\bf $\bullet$} H$_2$ is an important coolant of shocked gas, including the virialized gas that cools to form early galaxies.
 To fully capitalize on this potential, we need to harness both UV absorption from cool H$_2$, and rotational emission from warm H$_2$ for which models have already proven successful locally. This exciting science cannot be done with currently planned telescopes.   

{\bf $\bullet$} The excitation conditions of H$_2$ are strikingly variable in the local Universe,  encoding information on the accretion modality, heating conditions, AGN-driven feedback, and star-forming molecular reservoirs that is robust against the changing metal and dust abundance in the early Universe. Full exploitation of the information contained in the multiple H$_2$ cooling lines requires a very broad, sensitive, spectroscopic FIR wavelength coverage. 

{\bf $\bullet$} H$_2$ emission/absorption may be the only way to directly probe the gas cooling and feeding the most massive metal-free dark matter halos and to assess the molecular reservoirs inside dust- and metal-free star forming regions at the earliest epochs. This goal is within reach for lensed regions for the next generation of cryogenic FIR telescopes like OST.

\pagebreak

\end{document}